\documentclass{article}

\usepackage[T1]{fontenc}

\usepackage{geometry}
\usepackage{authblk}

\usepackage{subfig}

\usepackage{comment}
\usepackage[english]{babel}
\usepackage{xcolor}

\usepackage{graphicx}
\usepackage{latexsym}

\usepackage{amsthm}
\usepackage{mathtools}

\newtheorem{remark}{Remark}

\usepackage{amssymb}

\renewcommand{\appendix}{
\setcounter{section}{0}
\renewcommand{\thesection}{\Roman{section}}
\vspace{0.5cm}
{\Large{\bf APPENDIX}}}

\newcommand{\T}{\mathcal{T}}

\newcommand{\vv}{{\bf v}}

\usepackage{mathtools}

\title{Rebuttal of Morris' criticism of the diffusive compressible Euler model}
\author{Magnus Sv{\"a}rd}

  \affil{Dept. of Mathematics, University of Bergen, P.O. Box 7803, 5020 Bergen, Norway, \\ Magnus.Svard@uib.no}

\date{\today}
  
\begin{document}

\maketitle

\begin{abstract}
This short note addresses the criticism of the diffusive compressible Euler model regarding heat diffusion, sound attenuation and material frame indifference put forward by M. Morris.
\end{abstract}

\section{The model}

The diffusive compressible Euler (dcE) model (previously Eulerian model) is given by,
\begin{subequations}\label{eulerian} \\
\begin{align}
\partial_t \rho + \nabla\cdot(\rho \vv )&= \nabla\cdot  (\nu \nabla \rho),\label{continuity} \\
\partial_t (\rho \vv) + \nabla\cdot(\rho \vv \otimes \vv) + \nabla p&=\nabla\cdot 
(\nu \nabla \rho \vv), \qquad t\in [0,\T] \label{momentum} \\
\partial_t (E) + \nabla\cdot(E \vv +p\vv)  &= \nabla\cdot ( \nu \nabla E) + \nabla\cdot(\kappa_T \nabla T), \label{energy} \\ 
p&=\rho R T, \quad \textrm{ideal gas law,}\label{gaslaw}
\end{align}
\end{subequations}
where $\rho,\rho \vv, E,p,T$ are  density, momentum, total energy, pressure and temperature. Furthermore, $\vv=(u,v,w)$ is the velocity vector, $\gamma=c_p/c_v$ where $c_{p,v}$ are the heat capacities at constant pressure or volume.  For ideal gases $1<\gamma\leq 5/3$.  The total energy satisfies $E=\frac{p}{\gamma-1}+\frac{\rho |\vv|^2}{2}$. Furthermore, $R$ is the gas constant, $\nu=\nu(\rho,T)$ the diffusion coefficient and $\kappa_T$ the  heat diffusion coefficient. In the original model (\cite{Svard18}) $\kappa_T$ was not included, i.e., $\kappa_T=0$. In its current form the diffusive coefficients take the values (\cite{Svard24}),
\begin{align}
\nu =\frac{\mu}{\rho},\quad\quad \kappa_T = \frac{\mu c_p}{Pr}(1-Pr)\label{nu}
\end{align}
where $\mu$ is the value of the dynamic viscosity and $Pr$ is the Prandtl number.
\begin{remark}
Although $\mu$ takes the \emph{value} of the dynamic viscosity, the $\nu$ terms do not model viscosity. They model diffusion. 
\end{remark}
\begin{remark}
Note that $\nu$ and $\kappa_T$ in the dcE model, or $\mu$  and  $\kappa$ in the NSF model, are not fundamental constants of nature but {\bf model-specific parameters} associated with constitutive laws. These parameters can not be measured directly but the associated system must be solved (often approximately) for the experimental setup and the parameters optimised to make the (approximate) solutions match the available data. (See  \cite{Assael_etal23} for a discussion on heat conduction.)
\end{remark}
\begin{remark}
In order to prove the existence of weak solutions to (\ref{eulerian}), the $\nu$ and $\kappa_T$ had to be augmented with some extra terms (\cite{Svard22}). These terms may be considered technical assumptions and can be taken suitably small. The values given in (\ref{nu}) are the important ones with respect to modelling.
\end{remark}

Throughout, I make the distinction between diffusion, i.e., the transport by random movements, and conduction, i.e., the transfer of momentum and internal energy by collisions without molecules changing their relative positions. In gases, where the mean free path is far greater than the size of the molecules, diffusion is the dominating mechanism for random transport.

In \cite{Svard18}, it is stated that the model is derived for compressible and diffusive (viscous and heat conductive) ideal gases. It can be expected to be accurate for the same range of Knudsen numbers as the compressible Navier-Stokes-Fourier (NSF) system.

I have used ``viscous and heat conductive'' in the description of (\ref{eulerian}) to emphasise that it models the same types of gases as the viscous and heat conductive compressible Navier-Stokes equations. However, {\bf (\ref{eulerian}) does not model viscosity}. In fact, there is a long discussion on the stress tensor  and its role in the Navier-Stokes equations in \cite{Svard18}, that serves to emphasise the change of paradigms that subsequently leads to (\ref{eulerian}).

As the NSF system, the dcE model can for practical purposes be used effectively for gases that are close to ideal, such as air, \cite{DolejsiSvard21,SayyariDalcin21,SayyariDalcin21_2,SvardMunthe23,Svard24}. Under normal flow conditions the effects from a real gas law (e.g. van der Waal's),  or temperature dependent viscosity, heat conductive coefficients and specific heats are often small. In shocks, however, they may have significant effects and their inclusion will improve the results (\cite{SvardMunthe23}). In this respect, the dcE model is no different from the NSF.

\section{Initial remarks}

All models, new and old, should be subject to a critical examination. This sharpens arguments supporting the model; it may lead to improvements of the model and it highlights its limitations. This is what led me to consider alternatives to the NSF system and my proposition for a new model should of course be scrutinised in the same way.

Although, this is probably evident to the readership, I remind that scientific criticism should abide by the usual scientific rules. As the bare minimum, the most recent model/theory should be considered, misrepresentations avoided and range of validity acknowledged. One should also be careful not to extrapolate arguments and criticisms beyond their validity ranges.  Selective use of evidence should be avoided. Scientific disagreements should be acknowledged and  one side should not be presented as the consensus. 

I did not intend to respond to the stream of Morris' critical reports on ResearchGate. (That would require a substantial part of my research time.) I have already responded to most of the criticism regarding noble and polyatomic gases in \cite{SvardMunthe23,Svard24} and updated the model to include the $\kappa_T$ term. However, since there is now a critical paper published in a scientific journal (\cite{Morris24}), I am forced to address its inaccuracies. I also take the opportunity to address other recent re-iterations and extrapolations of Morris' previous arguments, as well as a new attempt to debunk the dcE model.

\section{Heat transfer}

In \cite{Morris24}, Morris claims that the dcE model ``substantially underestimates the magnitude of the heat flux in gases''.  To support this criticism of the dcE model,  Morris considers the, at the time superseded, model with $\kappa_T=0$. Already in \cite{SvardMunthe23}, it was tentatively proposed that the $\kappa_T$ term should be included. A physical rationale for its presence was given and a value was proposed for noble gases. In \cite{Svard23,Svard24}, the heat diffusion for the dcE model was studied and the correction was confirmed and generalised to non-noble gases. In its current form, the dcE system (\ref{eulerian}) models heat transfer as accurately as the NSF model for gases.

Furthermore, Morris goes on and claims that the dcE model fails to predict heat conduction in water. This criticism is irrelevant since the dcE system does not model liquids. (In liquids, the mean free path is of the same order as the size of the molecules. Hence, conduction, rather than diffusion, is the major mechanism for random transport.)

\section{Sound attenuation}

In \cite{Morris24_2}, the accuracy of (\ref{eulerian}) with respect to extremely high-frequency ultrasound experiments was criticised. Once again, Morris chooses, rather inexplicably, to include the superseded model with $\kappa_T=0$ along with the current model.

For the current model, with the diffusive coefficients given in (\ref{nu}), Morris finds that it reproduces the sound attenuation experiments for \emph{noble gases}, just as accurately as the NSF model does.

For real gases, $N_2,O_2$ and dry air, Morris observes a mismatch between the sound attenuation coefficient for the (linearised) dcE model and experimental values. Furthermore, Morris claims that the NSF system perfectly predicts the sound attenuation. This is achieved by \emph{choosing} the bulk viscosity in the NSF to obtain an exact match with the experimental sound attenuation coefficient. Since there is no parameter left to choose in the dcE model, this trick can not be done and therefore, according to Morris, the dcE model is incorrect.

This line of argumentation contains a number of statements and assumptions that requires further scrutiny.

{\bf Bulk viscosity:}  \emph{The bulk viscosity is not accurately known}, which is recognised by Morris in \cite{Morris24_2}.  A common choice for the bulk viscosity, that gives good agreement with data in aerodynamic experiments, is given by Stokes' hypothesis. For that choice of bulk viscosity, the NSF system is slightly \emph{worse} (about 6\% \cite{Svard24}) than the dcE model, for these extremely high-frequency ultrasound attenuation experiments.

  Whether or not Stokes' hypothesis is correct for diatomic gases has justifiably been questioned in the literature. Nevertheless, the choice made in \cite{Morris24_2} to demonstrate the superiority of the NSF over the dcE model can hardly  be considered to represent the current scientific consensus. Furthermore, the effects of this choice with respect to other, less extreme, validation experiments need to be evaluated.  

  {\bf Polyatomic gases:}  Morris terms $N_2,O_2,CH_4$ and air as, polyatomic \emph{ideal gases}. Here, one needs to be a bit more specific. Under normal conditions they are certainly very close to ideal. That is why both the NSF and dcE produce accurate aerodynamic results with ideal gas approximations of the fluid parameters. (Recall that in an ideal gas, molecules bounce elastically and the system does not exchange energy with the rotational or vibrational modes of the molecules.) 

  The paper \cite{BondChiang92} contains a discussion on the very high frequency sound experiments that Morris uses for validation of the linearised flow models. Here, I will draw the attention to a few statements. First, ``[the absorption] depends on the amplitude of the waves''. This was also noted in \cite{SvardMunthe23}, where very accurate numerical experiments were carried out. Only for extremely small amplitudes were the absorption coefficients calculated from the linearised equations recovered. For slightly higher amplitudes the actual absorption was higher than the linear prediction, due to non-linear effects in the full non-linear systems. (This applies both to the NSF and dcE models.) To allow for an estimate of the significance of this type of error, the amplitudes used in experiments must be known.

  Furthermore, Bond et al.  \cite{BondChiang92} identify the following contributions to the absorption coefficient  $\alpha=\alpha_{cl}+\alpha_{rot}+\alpha_{vib} $. $\alpha_{cl}$ is the coefficient obtained from classical theory, i.e., the linearised  NSF (or here the linearised dcE model). $\alpha_{rot}$ and $\alpha_{vib}$ are the contributions from the rotational and vibrational modes. If $\alpha_{rot}$ and $\alpha_{vib}$ are both close to zero, the gas behaves as an ideal gas. 

  However, with $f$ being the sound frequency and $p$ the ambient pressure, it is stated in \cite{BondChiang92} that for $f/p> 10 Hz/Pa$,  rotational relaxation phenomena may become important. That is, the gas is no longer ideal. The experiments Morris refers to are exceeding this limit. For monatomic gases, this is not a problem since they do not have rotational or vibrational modes that may store significant energy. Hence, the gas is essentially ideal and both the NSF and dcE accurately predict the absorption coefficient. For diatomic gases, the $f>1MHz$ experiments (at atmospheric pressure) excite rotational modes and the results deviate from the classical absorption coefficients. Furthermore, $CH_4$ has more internal modes that can store energy than $N_2$ and $O_2$, which distorts the coefficient further away from $\alpha_{cl}$. This explains why both the dcE model and the NSF model (without Morris' specific choice of bulk viscosity) underpredicts the actual attenuation coefficient for polyatomic gases. Both models are outside of their (close to) ideal gas regime. Notably, it is not suggested in \cite{BondChiang92} that the bulk viscosity in the NSF is chosen so as to compensate for the deviation from the measured absorption coefficient.

Nevertheless, bulk viscosity could be argued to model the energy interchange with rotational and vibrational modes in gases. If so, Morris' choice will indeed improve the models ability to reproduce sound attenuation at extremely high frequencies. However, this choice must then also be shown not to interfere with other aerodynamic validation cases, as well as attenuation at frequencies just below that where rotational modes are excited. 

  {\bf Liquids:} Morris is also using sound attenuation data for various liquids (water, mercury, benzene) in an attempt to falsify (\ref{eulerian}). I repeat: the system (\ref{eulerian}) is not a model for liquids.

  {\bf Summary:} Morris states that ``the dcE model fails to model sound wave attenuation in general fluids with physical accuracy'' (\cite{Morris24_2}). As I have shown above, this short statement is incorrect on at least two accounts. First, the dcE model is not valid for ``general fluids''. In particular, it is not valid for liquids; it is currently valid for essentially ideal gases. Second, contrary to Morris' claim, \emph{the dcE system does model sound attenuation as accurately as the NSF model when the gas behaves in a close to ideal way}. None of Morris' deliberations invalidate the dcE model under normal conditions. Not even with respect to sound attenuation for very high (but not extremely high) frequencies.

  In \cite{BondChiang92}, the complexity of ultra sound absorption for polyatomic gases (at very high to extreme frequencies) is discussed at length and there are many mechanisms beyond the classical (viscosity and heat conduction) that affect the absorption. 
  Hence, it seems difficult to justify that Morris' specific choice of bulk viscosity is the one that accounts for all the remaining frequency-dependent absorption mechanisms that influence the total absorption. This may be part of the reason why the bulk viscosity has not yet been decisively determined. 
  
  Nevertheless, even for these extreme frequencies, there is not a significant difference between the dcE model and standard versions of the NSF.

\begin{remark}
The update of the model in \cite{SvardMunthe23} was tentative since we had only investigated one data set (attenuation) and the experimental accuracy as well as the sound-wave amplitudes were not disclosed (\cite{Greenspan56}). (It was not common at the time of these experiments to report experimental errors.) With heat-diffusion data also pointing to the same correction (\cite{Svard24}) for noble gases, I could conclude that the model should be updated with the extra $\kappa_T$ term. Of course, this also means that the old sound attenuation measurements are accurate, but for polyatomic gases there is still an ambiguity regarding what physical phenomena they record.   
\end{remark}

\section{Material frame indifference}

In classical physics, Galilean invariance is a fundamental principle. It requires that the equations of motion  take the same form in any inertial frame.

Material frame indifference (MFI) is the notion that constitutive relations for materials take the same form in \emph{any} frame of reference.  The concept of MFI (or material objectivity) has been debated intensely for many decades. Not only if it is a generally valid physical principle, but also what it even means. See \cite{Frewer09} for a historical review. The review ends with a summary of the consensus that ``for solids and ordinary dense fluids the MFI-principle is a reasonable approximation to reduce constitutive equations.'' and ``This is no longer the case...if the mean free path length in the microprocesses gets longer as in gases and rarefied gases.'' Since (\ref{eulerian}) is a model for gases, it is thus far from obvious that its constitutive laws should satisfy the MFI-principle. 

Nevertheless, I will address the specific criticism put forward in \cite{Morris23}. There, Morris recasts the diffusive terms into something resembling viscous terms and proceeds to show that this construct is not invariant with respect to accelerating reference frames, and conclude that (\ref{eulerian}) is frame dependent. 

There is, however, a major logical flaw. \emph{Material indifference concerns constitutive laws that appear in the model. No constitutive law for viscous stresses is used in (\ref{eulerian}).} The model (\ref{eulerian}) is derived from conservation principles and {\bf  the parabolic terms in (\ref{eulerian}) model diffusion.}

Furthermore, if the actual constitutive laws are considered, i.e., the diffusive fluxes, they all take the same form as Fourier's law. Hence, and contrary to Morris' claim, \emph{the constitutive laws of (\ref{eulerian}) are indeed frame indifferent} (\cite{Wang99}).

\begin{remark}
One of the main points in the physical framework leading to (\ref{eulerian}) is the rejection of the view that gases should be viewed as a continuum of infinitesimal mass particles and that a force balance for the mass particles should be used to derive the governing equations. The mass particle view is what leads to the inclusion of a stress tensor in the NSF. In a gas, diffusion gives rise to viscous stresses, but in (\ref{eulerian}) diffusion is modelled directly which voids the need for the viscous stress tensor.
\end{remark}

\section{Final remarks}

In her most recent reports, Morris criticises
\begin{itemize}
\item the superseded dcE model despite an updated one being available.
\item the dcE system for not modelling liquids despite it being a model for gases.
\item the dcE system for not being frame indifferent with respect to some viscosity-resembling construct that is not a constitutive law in the model. (The actual constitutive laws in the dcE model are frame indifferent.)
\item the dcE system for not being accurate for extremely high frequency sound waves despite them being outside its (current) validity range. However, this was not enough to further the notion that the dcE model is flawed, since standard versions of the NSF can not accurately model such sound waves either. To be able to criticise the dcE model, Morris also had to tweak the NSF system such that it, ``by design'' as Morris puts it (\cite{Morris24_2}), is perfectly accurate.
\end{itemize}
In this note, I have addressed and rebutted Morris' criticism, and in doing so I also hope that I have shed some light on the methods that Morris uses in her attempts to falsify the dcE model.

The NSF system has been fine tuned for almost two centuries, by researchers who strived to improve the model and make it useful. Yet there are still aspects that are unclear, including the role and value of the bulk viscosity. (See \cite{Svard18} for other quirks with the NSF system.) From a computational/mathematical perspective, the NSF system is very challenging, and this is a consequence of the mass element approach in its derivation. A clear advantage of the dcE model is its weak well-posedness allowing it to be solved numerically much more efficiently and reliably.

Finally, the model (\ref{eulerian}) is work in progress and  the current version does not account for (strongly) non-ideal behaviour of the gas. That does not mean that it is impossible to augment the dcE system further to improve the modelling of non-ideal effects. However, the dcE model is not, and probably never will be,  a model for liquids, since the underlying physics is fundamentally different.


\end{document}